\documentclass{article}
\usepackage{spconf,amsmath,graphicx}
\usepackage{multirow}
\usepackage{amsfonts}
\usepackage{bm}
\usepackage{mathtools}
\usepackage{subcaption}

\DeclareMathOperator*{\argmax}{arg\,max}

\title{Joint speaker diarisation and tracking in switching state-space model}
\name{Jeremy H. M. Wong and Yifan Gong}
\address{Microsoft, USA}
\begin{document}
\ninept
\maketitle
\begin{abstract}

Speakers may move around while diarisation is being performed. When a microphone array is used, the instantaneous locations of where the sounds originated from can be estimated, and previous investigations have shown that such information can be complementary to speaker embeddings in the diarisation task. However, these approaches often assume that speakers are fairly stationary throughout a meeting. This paper relaxes this assumption, by proposing to explicitly track the movements of speakers while jointly performing diarisation within a unified model. A state-space model is proposed, where the hidden state expresses the identity of the current active speaker and the predicted locations of all speakers. The model is implemented as a particle filter. Experiments on a Microsoft rich meeting transcription task show that the proposed joint location tracking and diarisation approach is able to perform comparably with other methods that use location information.

\end{abstract}
\section{Introduction}

Speaker diarisation is the task of clustering segments of audio that are uttered by the same speaker. This can be used with speech recognition to provide rich transcriptions of audio, expressing both words and speaker identities. The task of diarisation can be broken down into counting the number of clusters and clustering the audio segments. By treating these sub-tasks separately, the number of clusters can first be estimated by finding the maximum gap in a chosen statistic \cite{tibshirani2001,park2020}, then the segments can be clustered using either $k$-means \cite{shum2011} or spectral clustering \cite{ning2006}. Alternatively, both sub-tasks can be performed in unison in the Agglomerative Hierarchical Clustering (AHC) framework \cite{siegler1997,jin1997}. This iteratively performs greedy merging of clusters based on a measured affinity, until a stopping criterion is reached. A Hidden Markov Model (HMM) can capture information about the temporal nature of speech, which may be useful for diarisation. The HMM can either be used within AHC in the computation of the affinities \cite{ajmera2003}, or on its own after being given an upper bound of the number of clusters \cite{diez2019,landini2020}.

Diarisation is often performed using only speaker embeddings, which are extracted using models that are trained to discriminate between speakers through a speaker identification or speaker verification task. Information about the locations of the speakers may be complementary to the speaker embeddings. Such location information is available when using a microphone array. In the HMM framework, location information can be incorporated by using the speaker embeddings together with either time-delay-of-arrival \cite{pardo2007,vijayasenan2012} or Sound Source Localisation (SSL) \cite{wong2021} features as the observations. In these works, the HMM state only encodes information about the identities of the speakers, and does not keep track of where each speaker is at each point in time. This therefore may not explicitly model the movements of speakers, and may assume that speakers are fairly stationary throughout a meeting.

In the vision domain, multi-face tracking can be achieved using separate Kalman filters to track the movements of each face \cite{shaik2007,foytik2011}. When using a microphone array, localisation information from the audio has been shown to be complementary to visual information for face tracking \cite{gebru2015}. In the audio-only scenario, challenges such as LOCATA \cite{evers2000} help to spur the development of audio localisation and tracking methods. Several of these methods also rely on Kalman or particle filtering techniques, to track the locations of a single \cite{bechler2003,salvati2018,markovic2010} or multiple \cite{segura2008} audio sources. When tracking multiple audio sources, multi-target extensions of probabilistic data association provide a framework to estimate which observations belong to each of the targets being tracked \cite{gehrig2006}. However, when used with multiple speakers \cite{murase2005,mcdonough2013,plinge2014}, these tracking methods often only rely on location information, and not speaker embeddings.

This paper proposes to track speaker movements jointly with performing diarisation, while also using speaker embeddings. It is hoped that explicitly modelling the movements of speakers may be beneficial to the diarisation task. A switching state-space model \cite{ghahramani2000} is proposed, that does joint modelling through a hidden state that encodes both information about the active speaker identity and also the current locations of each of the speakers. This model is implemented using a particle filter framework to accommodate for the forms of transition and emission likelihoods that are used. The model is referred to as the Switching State-space Particle Filter (SSPF).

\section{Joint clustering and location tracking}
\label{sec:sspf}

The HMM that is used for diarisation often encodes the current active speaker as the hidden state. In the work in \cite{wong2021}, the HMM computes the observation sequence likelihood as
\begin{equation}
p\left(\mathbf{D}_{1:T},\mathbf{S}_{1:T}\right)\approx\sum_{\mathbf{q}_{1:T}}\prod_{t=1}^Tp\left(\mathbf{d}_t\middle|q_t\right)p\left(\mathbf{s}_t\middle|q_t\right)p\left(q_t\middle|q_{t-1}\right),
\label{eq:hmm}
\end{equation}
where $\mathbf{d}_t$ and $\mathbf{s}_t$ are the speaker embedding and SSL features respectively at frame $t$, $T$ is the number of frames, and $q_t$ is the discrete hidden state that encodes the active speaker identity. The initial state probability is omitted here for brevity. In this formulation, it is not possible to infer where each speaker is at each point in time. Thus, the model does not explicitly capture the movements of speakers.

In order to track speaker movements, this paper proposes to encode the current active speaker identity as well as the current locations of all of the speakers in the hidden state. Furthermore, multiple concurrent active speakers are allowed, to accommodate overlapping speech. Speech separation is applied to the microphone array audio, forming $N$ channels without concurrent speakers in each. The SSPF simultaneously models all channels. In contrast, \cite{wong2021} merges the channels into a single stream. The SSPF hidden state is defined as
\begin{equation}
\mathbf{z}_t=\left\{\mathbf{q}_{t,1:N},\bm{\theta}_{t,1:M}\right\},
\end{equation}
where $q_{t,n}$ is a discrete variable representing the active speaker at frame $t$ in channel $n$, $\theta_{t,m}$ represents the angular location in radians around the microphone array at frame $t$ for speaker $m$, and $M$ is the number of speakers. Using the same Markov assumptions as \eqref{eq:hmm}, the observation sequence likelihood is computed as
\begin{align}
p\left(\mathbf{D}_{1:T,1:N},\mathbf{X}_{1:T,1:N}\right)\approx\sum_{\mathbf{Z}_{1:T}}\prod_{t=1}^T&\,p\left(\mathbf{D}_{t,1:N}\middle|\mathbf{z}_t\right)p\left(\mathbf{X}_{t,1:N}\middle|\mathbf{z}_t\right)\notag\\
&\times p\left(\mathbf{z}_t\middle|\mathbf{z}_{t-1}\right),
\label{eq:switching_state_observation_sequence_likelihood}
\end{align}
where $\mathbf{X}_{t,1:N}$ is used as a placeholder to represent a location-based observation feature that can take several possible forms. Here, diarisation is performed after speech separation, and thus each frame has $N$ unmixed observations, $\mathbf{D}_{t,1:N}$ and $\mathbf{X}_{t,1:N}$.

The transition probability is factorised for each state entity,
\begin{equation}
p\left(\mathbf{z}_t\middle|\mathbf{z}_{t-1}\right)\approx\left[\prod_{n=1}^NP\left(q_{t,n}\middle|q_{t-1,n}\right)\right]\!\left[\prod_{m=1}^Mp\left(\theta_{t,m}\middle|\theta_{t-1,m}\right)\right]\!\!.
\label{eq:factorised_transition}
\end{equation}
This assumes that each separate $q_{t,n}$ and $\theta_{t,m}$ propagate independently over time. The speaker transition probability, $P\left(q_{t,n}\middle|q_{t-1,n}\right)$ is an $M\times M$ matrix that is shared across all channels. The angular location transition likelihood is chosen to be a von Mises density function that is shared across all speakers,
\begin{equation}
p\left(\theta_{t,m}\middle|\theta_{t-1,m}\right)=\frac{1}{2\pi I_0\left(\varsigma\right)}e^{\varsigma\cos\left(\theta_{t,m}-\theta_{t-1,m}\right)},
\end{equation}
where the concentration, $\varsigma$, expresses how fast speakers tend to move, and $I_\nu\left(\varsigma\right)$ is the modified Bessel function of the first kind with order $\nu$. The von Mises density function is chosen to abide by $\theta_{t,m}$ being bounded by $\left(-\pi,\pi\right]$ with a periodic boundary condition.

The initial state likelihood, which is omitted in \eqref{eq:switching_state_observation_sequence_likelihood} for brevity, is similarly factorised into each of the separate state entities,
\begin{equation}
p\left(\mathbf{z}_1\right)\approx\left[\prod_{n=1}^NP\left(q_{1,n}\right)\right]\left[\prod_{m=1}^Mp\left(\theta_{1,m}\right)\right].
\label{eq:factorised_initial_state}
\end{equation}
Both the active speaker initial state probability, $P\left(q_{1,n}\right)$, and the initial location likelihood, $p\left(\theta_{1,m}\right)$, are set to be uniform, because the model has no information about the identity of the active speaker or the locations of the speakers, before any observation is made.

Similarly, the speaker embedding emission likelihood is also factorised into separate channels,
\begin{equation}
p\left(\mathbf{D}_{t,1:N}\middle|\mathbf{z}_t\right)\approx\prod_{n=1}^Np\left(\mathbf{d}_{t,n}\middle|\mathbf{z}_t\right),
\end{equation}
which makes the assumption that the emissions of the channels are independent of each other when given the state. Similarly to \cite{wong2021}, the emission likelihood for each channel is chosen to be a von Mises-Fisher density function,
\begin{equation}
p\left(\mathbf{d}_{t,n}\middle|\mathbf{z}_t\right)=\frac{\gamma^{\frac{\mathbb{D}}{2}-1}}{\left(2\pi\right)^{\frac{\mathbb{D}}{2}}I_{\frac{\mathbb{D}}{2}-1}\left(\gamma\right)}e^{\gamma\bm{\mu}_{q_{t,n}}\cdot\mathbf{d}_{t,n}},
\end{equation}
where $\mathbb{D}$ is the speaker embedding dimension, $\bm{\mu}_{q_{t,n}}$ represents the embedding centroid for speaker $q_{t,n}$, and $\gamma$ is the concentration. The log-likelihood is a cosine similarity between $\bm{\mu}_{q_{t,n}}$ and $\mathbf{d}_{t,n}$.

The location emission likelihood is also factorised per-channel,
\begin{equation}
p\left(\mathbf{X}_{t,1:N}\middle|\mathbf{z}_t\right)\approx\prod_{n=1}^Np\left(\mathbf{x}_{t,n}\middle|\mathbf{z}_t\right),
\label{eq:factorise_location_emission}
\end{equation}
which again makes the assumption that the observed locations in each channel are independent of each other when given the current state. Two forms of location features are considered. The first is the SSL vector, $\mathbf{s}_{t,n}$, which represents a categorical distribution, where each dimension expresses the probability that the sound had originated from each angular bin around the microphone array,
\begin{equation}
s_{t,n,i}=P\left(\psi=i\middle|\mathbf{x}_{t,n}\right),
\end{equation}
where $i$ is the angular bin index and $\psi$ is the angular bin from which the frame $\mathbf{x}_{t,n}$ may have originated. This is computed using a complex angular central Gaussian model \cite{ito2016}, as is described in \cite{yoshioka2019b}. The second form of location feature is the Direction-Of-Arrival (DOA), $\phi_{t,n}$, which is computed as the mode of the SSL,
\begin{equation}
\phi_{t,n}=b_j\quad ,\text{ where}\quad j=\argmax_is_{t,n,i},
\end{equation}
and $b_j$ is the angle in radians of the $j$th bin. An alternative is to compute the DOA as the circular mean of the SSL, similarly to \eqref{eq:circular_mean}, instead of the mode, but initial tests did not suggest any significant performance difference between the two choices.

When using the DOA as the observed location feature, $\mathbf{x}_{t,n}$ is substituted with $\phi_{t,n}$ in \eqref{eq:factorise_location_emission}, and the location emission likelihood for each channel can be computed as a von Mises density function,
\begin{equation}
p\left(\phi_{t,n}\middle|\mathbf{z}_t\right)=\frac{1}{2\pi I_0\left(\kappa\right)}e^{\kappa\cos\left(\phi_{t,n}-\theta_{t,q_{t,n}}\right)},
\label{eq:doa_emission}
\end{equation}
where the concentration, $\kappa$, expresses the observation noise. This measures a similarity between the observed location, $\phi_{t,n}$, and the predicted location of the speaker that is estimated to be active on the channel, $\theta_{t,q_{t,n}}$.

Alternatively, the full SSL vector can be used as the observed location feature, by substituting $\mathbf{x}_{t,n}$ with $\mathbf{s}_{t,n}$ in \eqref{eq:factorise_location_emission}. For this feature, the location emission likelihood for each channel is computed using a continuous categorical density function \cite{gordonrodriguez2020},
\begin{equation}
p\left(\mathbf{s}_{t,n}\middle|\mathbf{z}_t\right)=\frac{1}{C\left(\bm{\lambda}_{t,n}\right)}\prod_{i=1}^\mathbb{S}\lambda_{t,n,i}^{s_{t,n,i}},
\label{eq:continuous_categorical}
\end{equation}
where $\mathbb{S}$ is the number of discrete angular bins and $C\left(\bm{\lambda}_{t,n}\right)$ is the normalisation term defined in \cite{gordonrodriguez2020}. The continuous categorical bin probabilities are computed as a discretised von Mises distribution about a mean that represents the predicted location, $\theta_{t,q_{t,n}}$, of the current active speaker in the channel,
\begin{equation}
\lambda_{t,n,i}=\frac{e^{\kappa\cos\left(b_i-\theta_{t,q_{t,n}}\right)}}{\sum\limits_{j=1}^\mathbb{S}e^{\kappa\cos\left(b_j-\theta_{t,q_{t,n}}\right)}}.
\label{eq:discrete_vm}
\end{equation}
The equivalent log-likelihood of \eqref{eq:continuous_categorical} is a KL-divergence between the predicted SSL, $\bm{\lambda}_{t,n}$, and the measured SSL, $\mathbf{s}_{t,n}$, both of which represent discrete categorical distributions. Substituting \eqref{eq:discrete_vm} into \eqref{eq:continuous_categorical} yields
\begin{equation}
p\left(\mathbf{s}_{t,n}\middle|\mathbf{z}_t\right)=\frac{e^{\rho_{t,n}\cos\left(\eta_{t,n}-\theta_{t,q_{t,n}}\right)}}{C\left(\bm{\lambda}_{t,n}\right)\sum\limits_{j=1}^\mathbb{S}e^{\kappa\cos\left(b_j-\theta_{t,q_{t,n}}\right)}},
\label{eq:exact_ssl_emission}
\end{equation}
where
\begin{equation}
\rho_{t,n}=\kappa\sqrt{\sum_{i=1}^\mathbb{S}\sum_{j=1}^\mathbb{S}s_{t,n,i}s_{t,n,j}\cos\left(b_i-b_j\right)}
\end{equation}
and
\begin{equation}
\eta_{t,n}=\tan^{-1}\left(\frac{\sum\limits_{i=1}^\mathbb{S}s_{t,n,i}\sin b_i}{\sum\limits_{i=1}^\mathbb{S}s_{t,n,i}\cos b_i}\right).
\label{eq:circular_mean}
\end{equation}
This suggests that with the choice of location emission likelihood of \eqref{eq:continuous_categorical} and \eqref{eq:discrete_vm}, the SSL, $\mathbf{s}_{t,n}$, at each frame can be completely characterised by an equivalent concentration, $\rho_{t,n}$, and mean, $\eta_{t,n}$. The concentration may weigh the contribution of each frame to the total log-likelihood proportionally to the sharpness of the SSL.

\begin{figure}[t]
\centering
\begin{subfigure}[b]{0.232\textwidth}
\centering
\includegraphics[width=\textwidth]{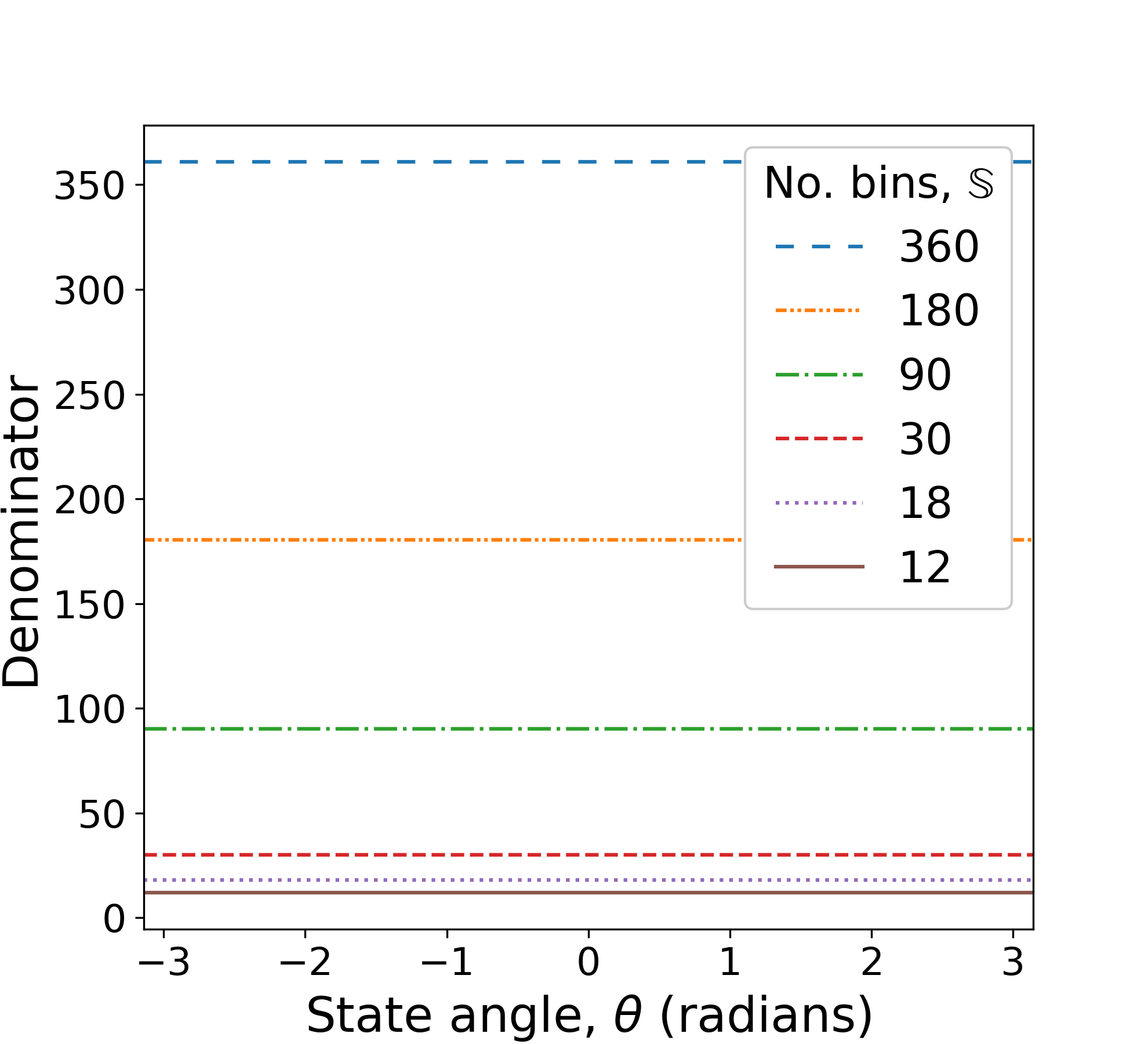}
\caption{$\kappa =0.1$}
\end{subfigure}
\begin{subfigure}[b]{0.232\textwidth}
\centering
\includegraphics[width=\textwidth]{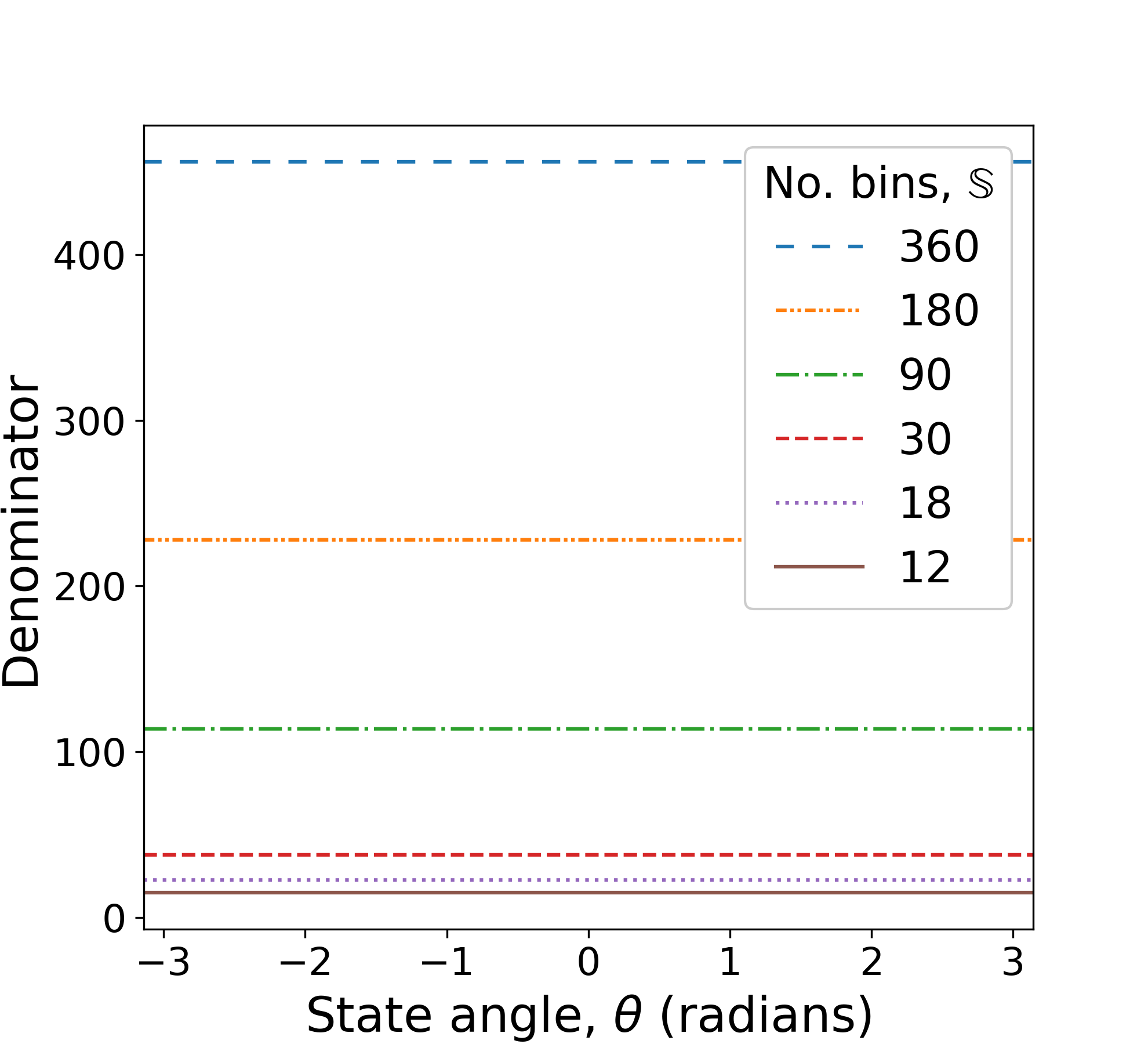}
\caption{$\kappa =1$}
\end{subfigure}
\begin{subfigure}[b]{0.232\textwidth}
\centering
\includegraphics[width=\textwidth]{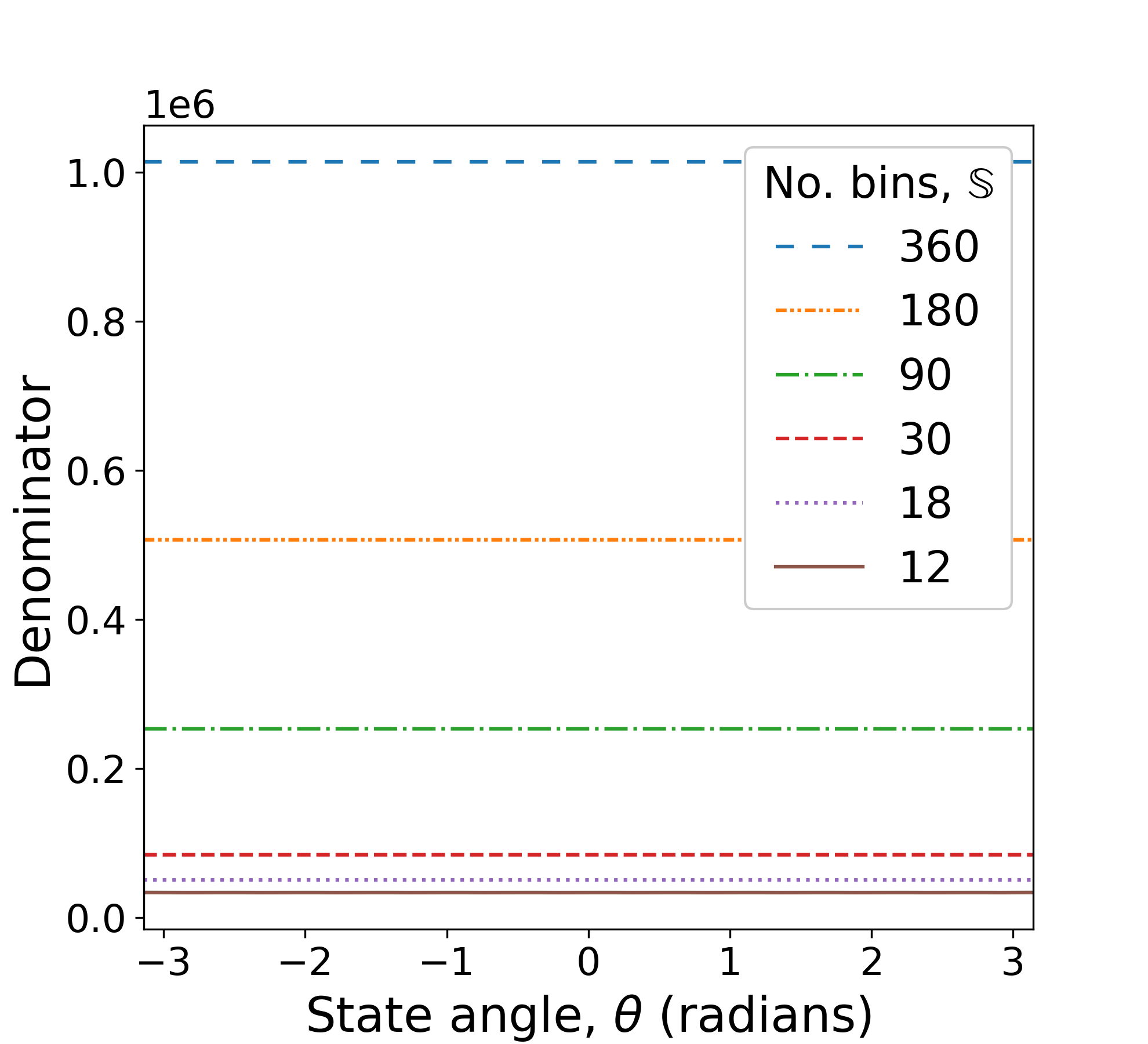}
\caption{$\kappa =10$}
\end{subfigure}
\begin{subfigure}[b]{0.232\textwidth}
\centering
\includegraphics[width=\textwidth]{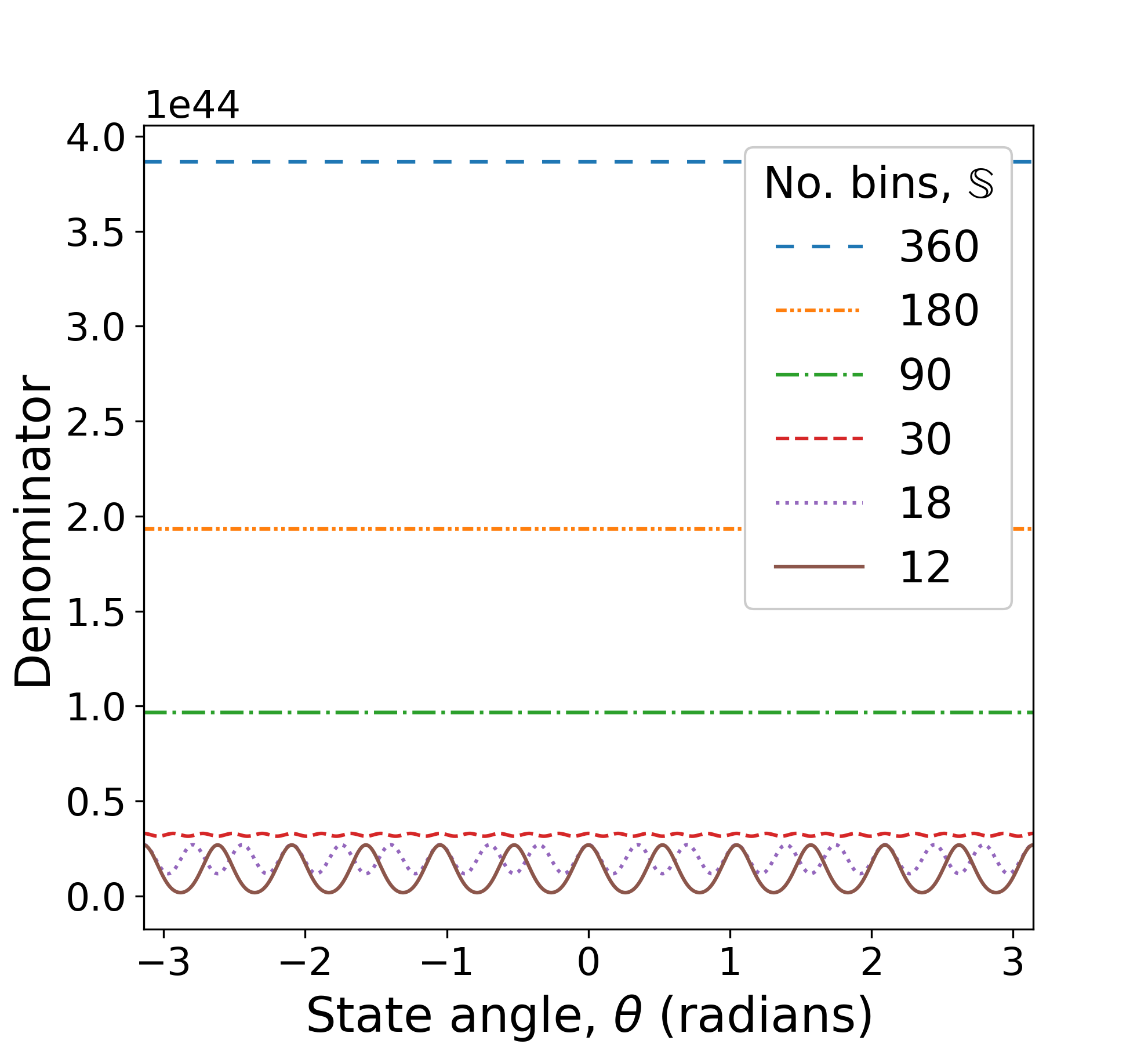}
\caption{$\kappa =100$}
\end{subfigure}
\caption{Denominator term of discretised von Mises distribution \eqref{eq:discrete_vm}}
\label{fig:denominator}
\end{figure}

However, the normalisation term, $C\left(\bm{\lambda}_{t,n}\right)$, is difficult to compute in a numerically stable manner \cite{gordonrodriguez2020}. As such, it is ignored in this paper. Furthermore, the $\sum_je^{\kappa\cos\left(b_j-\theta_{t,q_{t,n}}\right)}$ term in the denominator of \eqref{eq:exact_ssl_emission} is also ignored. Figure \ref{fig:denominator} plots $\sum_je^{\kappa\cos\left(b_j-\theta\right)}$ as a function of $\theta$, for various values of $\kappa$ and $\mathbb{S}$. The plots suggest that $\sum_je^{\kappa\cos\left(b_j-\theta\right)}$ is approximately independent of $\theta$, except when both the concentration, $\kappa$, is large and the number of angular bins, $\mathbb{S}$, is small. The setup in this paper does not operate in this regime. Thus it seems reasonable to omit this term. Therefore, the location emission likelihood is computed as
\begin{equation}
p\left(\mathbf{s}_{t,n}\middle|\mathbf{z}_t\right)\propto e^{\rho_{t,n}\cos\left(\eta_{t,n}-\theta_{t,q_{t,n}}\right)},
\label{eq:ssl_emission}
\end{equation}
which looks similar in form to a von Mises density function.

With $N$ separated channels, there may not be $N$ concurrent active speakers at every frame. If frame $t$ in channel $n$ does not have an observation, then the emission likelihoods are set to $p\left(\mathbf{d}_{t,n}\middle|\mathbf{z}_t\right)=1$ and $p\left(\mathbf{x}_{t,n}\middle|\mathbf{z}_t\right)=1$ for this frame and channel.

\begin{figure}[t]
\centering
\includegraphics[width=0.48\textwidth]{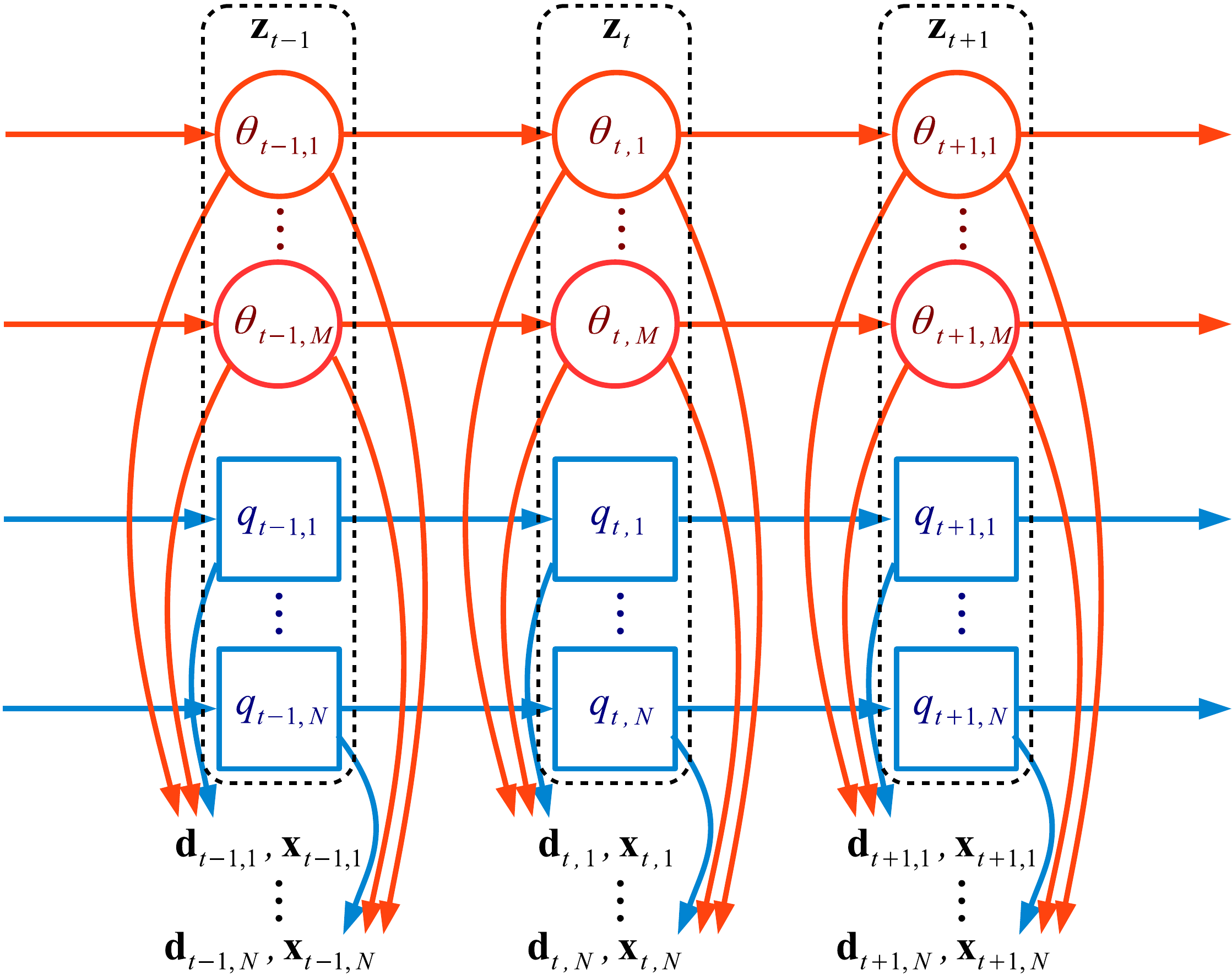}
\caption{Joint modelling of discrete speaker turns (squares) and continuous locations (circles) using a switching state-space model}
\label{fig:graphical_model}
\end{figure}

The joint speaker turn and location tracking model is illustrated graphically in Figure \ref{fig:graphical_model}. This is reminiscent of the switching state-space model proposed in \cite{ghahramani2000}. The $N$ discrete chains that express the current active speakers switch the outputs between the $M$ continuous chains that track the locations of each of the speakers, to generate the observed locations. The parameters of the model are the speaker embedding centroids, $\bm{\mu}_{1:M}$, the speaker transition probabilities, $P\left(q_{t,n}\middle|q_{t-1,n}\right)$, and the concentrations, $\gamma$, $\varsigma$, and $\kappa$. The concentrations are estimated using parameter sweeps on the $\emph{dev}$ data, while the speaker embedding centroids and speaker transition probabilities are maximum likelihood estimates from the hypothesised clusters from an initial AHC run. Uniform smoothing is interpolated into the speaker transition probabilities to improve generalisation.

\section{Particle filter implementation}

Performing clustering by decoding the model requires computing the forward recursion of
\begin{align}
p\left(\mathbf{z}_t\middle|\mathbf{O}_{1:t,1:N}\right)=&\int p\left(\mathbf{z}_{t-1}\middle|\mathbf{O}_{1:t-1,1:N}\right)p\left(\mathbf{D}_{t,1:N}\middle|\mathbf{z}_t\right)\notag\\
&\times p\left(\mathbf{X}_{t,1:N}\middle|\mathbf{z}_t\right)p\left(\mathbf{z}_t\middle|\mathbf{z}_{t-1}\right)d\mathbf{z}_{t-1},
\label{eq:forward}
\end{align}
where $\mathbf{O}_{1:t,1:N}$ is used to concisely represent the pair of observations, $\left\{\mathbf{D}_{1:t,1:N},\mathbf{X}_{1:t,1:N}\right\}$. The choice of emission and transition likelihoods in Section \ref{sec:sspf} are not closed under the multiplication and convolution operations in \eqref{eq:forward}. This makes it difficult to implement the model exactly, in a manner analogous to a Kalman filter or HMM. In this paper, the model is implemented as a particle filter \cite{gordon1993}. This does not require the likelihoods to be closed under the forward pass operations, and instead performs a Monte Carlo simulation of the propagation of density functions along the forward pass.

The sequential importance resampling algorithm \cite{rubin1987} is used. At each frame in the forward pass, the prediction step samples particles from either the initial state likelihood, $P\left(\mathbf{z}_1\right)$, for the first frame or from the transition likelihood, $P\left(\mathbf{z}_t\middle|\mathbf{z}_{t-1}\right)$, at subsequent frames, when given the particles or resampled particles from the previous frame. The factorised forms of \eqref{eq:factorised_transition} and \eqref{eq:factorised_initial_state} allow each state entity to be sampled separately. The collection of particles represents an approximation of the prediction likelihood, 
\begin{equation}
p\left(\mathbf{z}_t\middle|\mathbf{O}_{1:t-1,1:N}\right)\approx\sum_{r=1}^\mathbb{R}\check{\omega}_{t-1}^{\left(r\right)}\delta\left(\mathbf{z}_t,\widehat{\mathbf{z}}_t^{\left(r\right)}\right),
\end{equation}
where $\widehat{\mathbf{z}}_t^{\left(r\right)}$ is the $r$th particle, $\mathbb{R}$ is the number of particles, $\check{\omega}_{t-1}^{\left(r\right)}$ are the importance weights after resampling from the previous frame, and the Dirac delta function is defined as
\begin{equation}
\delta\left(\mathbf{y}_1,\mathbf{y}_2\right)=\left\{\begin{matrix*}[l]\infty&,\text{ if }\mathbf{y}_1=\mathbf{y}_2\\0&,\text{ otherwise}\end{matrix*}\right..
\end{equation}

After sampling the particles, the update step then computes the importance sampling weights as
\begin{equation}
\omega_t^{\left(r\right)}=\frac{\check{\omega}_{t-1}^{\left(r\right)}p\left(\mathbf{D}_{t,1:N}\middle|\widehat{\mathbf{z}}_t^{\left(r\right)}\right)p\left(\mathbf{X}_{t,1:N}\middle|\widehat{\mathbf{z}}_t^{\left(r\right)}\right)}{\sum\limits_{r^\prime=1}^\mathbb{R}\check{\omega}_{t-1}^{\left(r^\prime\right)}p\left(\mathbf{D}_{t,1:N}\middle|\widehat{\mathbf{z}}_t^{\left(r^\prime\right)}\right)p\left(\mathbf{X}_{t,1:N}\middle|\widehat{\mathbf{z}}_t^{\left(r^\prime\right)}\right)}.
\label{eq:importance_weight}
\end{equation}
The collection of particles and importance weights now approximate the update likelihood,
\begin{equation}
p\left(\mathbf{z}_t\middle|\mathbf{O}_{1:t,1:N}\right)\approx\sum_{r=1}^\mathbb{R}\omega_t^{\left(r\right)}\delta\left(\mathbf{z}_t,\widehat{\mathbf{z}}_t^{\left(r\right)}\right).
\end{equation}

Often, sequential Monte Carlo simulation methods suffer from the importance weights attenuating to zero for many particles, as the forward pass progresses. This is because the importance weights are computed recursively as a product of previous importance weights in \eqref{eq:importance_weight}. This may make it difficult to effectively explore the support of the state space. Resampling \cite{rubin1987} aims to alleviate this at the expense of an increase in the variance of the estimates. A new collection of resampled particles are sampled with replacement from the original particles, $\widehat{\mathbf{z}}_t^{\left(r\right)}$, with each original particle being resampled with a probability equal to its importance weight, $\omega_t^{\left(r\right)}$. The systematic method \cite{carpenter1999} is used in this paper to perform resampling. After resampling, the new resampled importance weights are set uniformly, $\check{\omega}_t^{\left(r\right)}=\frac{1}{\mathbb{R}}$. Resampling is only performed at a frame if the effective sample size \cite{kong1994}, $\left[\sum_r\omega_t^{\left(r\right)2}\right]^{-1}$, falls below a threshold.

\section{Decoding}
\label{sec:decoding}

Clustering can be performed by decoding the model. Only the active speakers, $\mathbf{q}_{t,1:N}$, are of interest to the diarisation task, while the speaker locations, $\bm{\theta}_{t,1:M}$, can be marginalised over. One approach to estimate the active speaker sequence is to use a Viterbi-style decoding
\begin{equation}
\mathbf{Q}_{1:T,1:N}^*\!=\!\argmax_{\mathbf{Q}_{1:T,1:N}}\!\int\! p\left(\mathbf{O}_{1:T,1:N},\mathbf{Q}_{1:T,1:N},\bm{\theta}_{1:T,1:M}\right)d\bm{\theta}_{1:T,1:M}\!.
\end{equation}
However, it may not be trivial to develop an efficient algorithm for this when the hidden state contains continuous variables. Furthermore, in the diarisation setup used in this paper, the objective is to hypothesise a speaker identity for each word, which may not be perfectly matched with finding the most likely sequence.

Decoding is instead performed by first computing the per-frame speaker state posteriors, marginalising over the location states,
\begin{equation}
P\!\left(\mathbf{q}_{t,1:N}\middle|\mathbf{O}_{1:T,1:N}\right)\!=\!\!\!\int\!\! p\!\left(\mathbf{q}_{t,1:N},\bm{\theta}_{t,1:M}\middle|\mathbf{O}_{1:T,1:N}\right)d\bm{\theta}_{t,1:M}\!.
\label{eq:marginalise_location}
\end{equation}
The speaker for each word is then estimated by choosing the most probable speaker from the aggregated speaker state posteriors over the frames within the word. Aggregation of the state posteriors can be done either as a sum,
\begin{equation}
\overline{q}_l^*=\argmax_{\overline{q}_l}\sum_{t=\tau_l^\text{start}}^{\tau_l^\text{end}}P\left(q_{t,n_l}=\overline{q}_l\middle|\mathbf{O}_{1:T,1:N}\right),
\label{eq:sum_combination}
\end{equation}
a product,
\begin{equation}
\overline{q}_l^*=\argmax_{\overline{q}_l}\prod_{t=\tau_l^\text{start}}^{\tau_l^\text{end}}P\left(q_{t,n_l}=\overline{q}_l\middle|\mathbf{O}_{1:T,1:N}\right),
\label{eq:product_combination}
\end{equation}
or majority voting,
\begin{equation}
\overline{q}_l^*=\argmax_{\overline{q}_l}\!\!\sum_{t=\tau_l^\text{start}}^{\tau_l^\text{end}}\!\!\partial\left[\overline{q}_l,\argmax_{q}P\left(q_{t,n_l}=q\middle|\mathbf{O}_{1:T,1:N}\right)\right]\!\!,
\label{eq:max_combination}
\end{equation}
where $\overline{q}_l$ is the speaker identity of the $l$th hypothesised word, $\tau_l^\text{start}$ and $\tau_l^\text{end}$ are the start and end frame indexes of the word respectively, $n_l$ is the channel on which the word is detected, the Kronecker delta function is defined as
\begin{equation}
\partial\left(i,j\right)=\left\{\begin{matrix*}[l]1&,\text{ if }i=j\\0&,\text{ otherwise}\end{matrix*}\right.,
\end{equation}
and $P\left(q_{t,n_l}\middle|\mathbf{O}_{1:T,1:N}\right)$ is computed by marginalising over the other channels in $P\left(\mathbf{q}_{t,1:N}\middle|\mathbf{O}_{1:T,1:N}\right)$. The product combination in \eqref{eq:product_combination} is most closely related to a maximum probability interpretation, as the probability for the speaker of a word should be computed as a joint probability of the same speaker over all of the frames within the word.

The state posterior in \eqref{eq:marginalise_location} can be estimated through Forward Filtering-Backward Smoothing (FFBS) \cite{doucet2000},
\begin{equation}
p\left(\mathbf{q}_{t,1:N},\bm{\theta}_{t,1:M}\middle|\mathbf{O}_{1:T,1:N}\right)\approx\sum_{r=1}^\mathbb{R}\dot{\omega}_t^{\left(r\right)}\delta\left(\mathbf{z}_t,\widehat{\mathbf{z}}_t^{\left(r\right)}\right),
\end{equation}
where the backward recursion computes the backward importance weights as
\begin{equation}
\dot{\omega}_t^{\left(r\right)}=\omega_t^{\left(r\right)}\sum_{i=1}^\mathbb{R}\dot{\omega}_{t+1}^{\left(i\right)}\frac{p\left(\mathbf{\widehat{z}}_{t+1}^{\left(i\right)}\middle|\mathbf{\widehat{z}}_{t}^{\left(r\right)}\right)}{\sum\limits_{j=1}^\mathbb{R}\omega_t^{\left(j\right)}p\left(\mathbf{\widehat{z}}_{t+1}^{\left(i\right)}\middle|\mathbf{\widehat{z}}_{t}^{\left(j\right)}\right)}.
\label{eq:backward_importance_weight}
\end{equation}

In this paper, an exact computation of the backward importance weights in \eqref{eq:backward_importance_weight} is used, which has a computational cost that scales as $\mathcal{O}\left(\mathbb{R}^2\right)$. This can be expensive when using many particles. Many particles may be required to sufficiently explore the state space. A kernel density approximation \cite{friedman1977,gray2000} can be used to speed up the computation to scale as $\mathcal{O}\left(\mathbb{R}\log\mathbb{R}\right)$, but this requires that the transition likelihoods represent monotonic kernels \cite{klaas2006}, which may limit the form of the allowed active speaker transition probabilities, $P\left(q_{t,n}\middle|q_{t-1,n}\right)$, to matrices with a probability attenuating monotonically away from the diagonal. As opposed to this, the forward recursion has a computational cost that scales as $\mathcal{O}\left(\mathbb{R}\right)$. Therefore, the computational cost can be reduced by decoding using only the forward pass, by replacing $\mathbf{O}_{1:T,1:N}$ with $\mathbf{O}_{1:t,1:N}$ in the conditional dependencies in \eqref{eq:marginalise_location}, \eqref{eq:sum_combination}, \eqref{eq:product_combination}, and \eqref{eq:max_combination}. However, this foregoes information from the future context when making the decoding decisions.

An alternative method to reduce the computational cost is to uniformly sub-sample the particles after the forward pass, when performing the backward pass. The exploration of the state space in the FFBS algorithm is primarily achieved during the sampling of particles in the prediction step of the forward pass. Therefore, having a large number of particles is more important for the forward pass than the backward pass.

Decoding for diarisation is done per word. Thus, it seems reasonable to restrict the state transitions to only allow speaker changes at the word boundaries. In the forward pass, this can be achieved by setting $P\left(q_{t,n}\middle|q_{t-1,n}\right)$ to the identity matrix when sampling in the prediction step at frames that are not at word boundaries. In the backward pass, the same restricted speaker transition probabilities can be used to compute the backward importance weights in \eqref{eq:backward_importance_weight}.

\section{Meeting transcription setup}

The proposed approach was evaluated on a rich meeting transcription task, with the setup that was initially described in \cite{yoshioka2019b}, and used again in \cite{wong2021}. Audio from a microphone array was separated into multiple channels, with the assumption that there were no concurrent speakers within each channel. Voice activity detection and speech recognition were run on each channel. Speaker change detection was used to find segments with speaker purity, by applying a threshold to the cosine similarity of the speaker embeddings computed using the model described in \cite{zhou2021}. AHC was then used to cluster together all of the segments from all of the channels that belonged to the same speaker, by greedily merging clusters with the highest speaker embedding cosine similarity, until the maximum similarity fell below a threshold. The Hungarian algorithm was then used to find the optimal mapping between the AHC hypothesised clusters and the enrolled speakers. These tagged AHC clusters were used to initialise the parameters of either a HMM or SSPF model, which then refined the clusters. The maximum number of active speakers, $M$, was equal to the number of AHC clusters. As with in \cite{wong2021}, the HMM parameters here were fine-tuned for each meeting using expectation-maximisation. The SSPF parameters were not modified after initialisation. In \cite{wong2021}, Hungarian speaker tagging was performed after HMM clustering. However, in this paper, HMM or SSPF clustering was performed after Hungarian tagging, to isolate the experimental trends associated with the HMM and SSPF methods, and ignore the trends due to the interactions between clustering and tagging. Following \cite{wong2021}, the HMM here also used a segment of one or more words as a frame. A uniform time segmentation may be essential to effectively model the temporal movements of speakers in the SSPF. As such, the SSPF used frames with a duration and shift of 0.4s.

\section{Experiments}

Audio data was collected from internal Microsoft meetings, with an average of 7 active participants per meeting, lasting up to 1 hour each. The \emph{dev} set comprised 51 meetings making up 23 hours, while the \emph{eval} set comprised 60 meetings making up 35 hours. The model described in \cite{zhou2021} was used to extract 128-dimensional $d$-vector speaker embeddings. The dimension of the SSL vectors was 360. The baseline HMM used SSL vectors that were downsampled to 18 dimensions, as this yielded improvements in initial tests. The SSPF used the full 360-dimensional SSL vectors, to retain the spatial resolution for accurate location tracking. The speaker-attributed Word Error Rate (WER) \cite{yoshioka2019b} was used to measure the performance. This was computed by measuring the WER separately for each speaker, then averaging the WERs over all speakers. The speaker-attributed WER assesses both the speaker diarisation and speech recognition performances together, which are both important for the rich meeting transcription task.

\begin{figure}[t]
\centering
\begin{subfigure}[b]{0.48\textwidth}
\centering
\includegraphics[width=\textwidth]{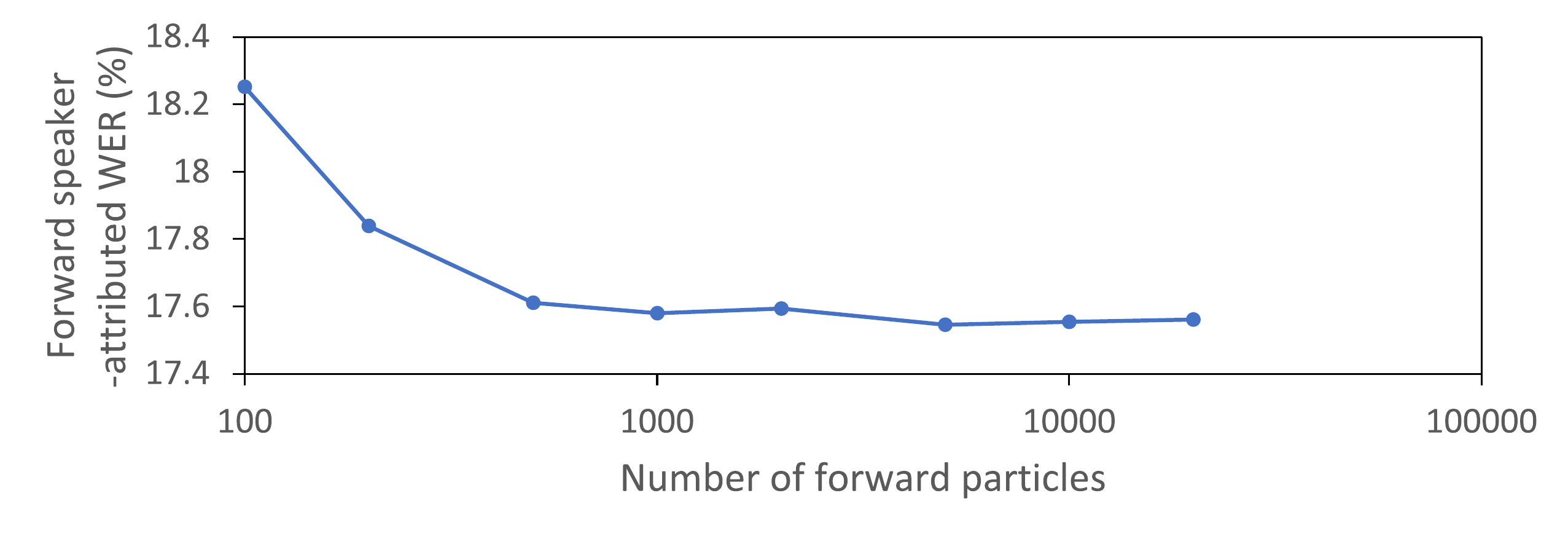}
\caption{Number of forward particles}
\label{fig:num_forward_particles}
\end{subfigure}
\begin{subfigure}[b]{0.48\textwidth}
\centering
\includegraphics[width=\textwidth]{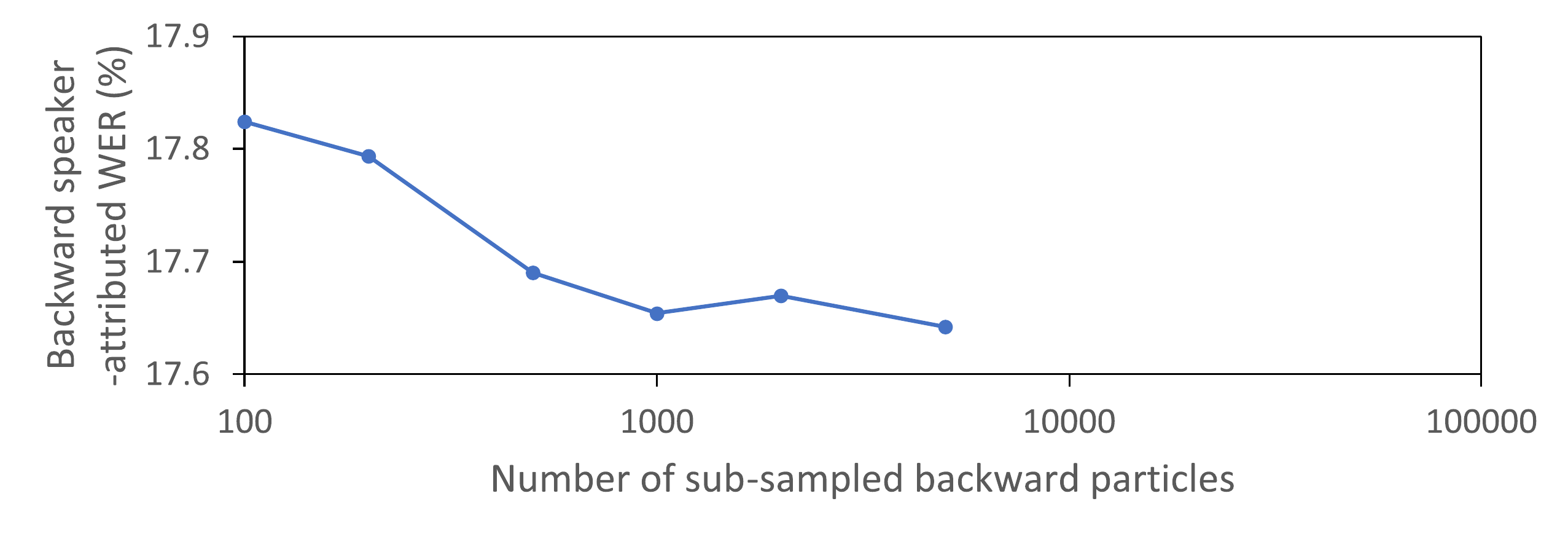}
\caption{Number of backward particles sub-sampled from 20000 forward particles}
\label{fig:num_backward_particles}
\end{subfigure}
\caption{Performance on the \emph{dev} set with various numbers of forward and sub-sampled backward particles}
\label{fig:num_particles}
\end{figure}

The first experiment assesses the influence of the number of particles. This primarily affects the exploration of the state space during the forward pass. As is explained in Section \ref{sec:decoding}, the particles can be sub-sampled during the backward pass to reduce the computational cost. Decoding of the SSPF can be done using only a forward pass or by using both the forward and backward passes. Figure \ref{fig:num_forward_particles} assesses the impact on the \emph{dev} set of the number of particles in the forward pass, by decoding using only the forward pass. DOA features were used with sum aggregation, without state transition restrictions. The speaker-attributed WER can be seen to degrade when fewer than 1000 particles are used. It may not be possible to effectively explore the support of the state space with so few particles. In the remaining experiments, 20000 particles were used in the forward pass to ensure adequate exploration of the state space. Going beyond 20000 particles required more than the available CPU memory, as this implementation was not optimised for memory efficiency.

Decoding using only the forward pass ignores information about the future context. Such information can be utilised by performing decoding using FFBS. In the backward pass, the computational cost can be reduced by sub-sampling the particles from the 20000 in the forward pass. Figure \ref{fig:num_backward_particles} assesses how the number of sub-sampled particles used in the backward pass affects the performance on the \emph{dev} set. The speaker-attributed WER improves as more sub-sampled particles are used. As a comparison between the two passes, a forward pass with 20000 particles yielded a speaker-attributed WER of 17.56\%, while a backward pass with 5000 sub-sampled particles yielded 17.64\%. It is a reasonable guess that the performance of the backward pass may eventually surpass that of the forward pass when given sufficient sub-sampled particles. However, going beyond 5000 sub-sampled particles required infeasible computation times in the current implementation. Unless otherwise stated, the remaining experiments perform decoding using only the forward pass.

The next experiment investigates the benefit of tracking the speaker locations, for the diarisation task. The SSPF model can use only speaker embeddings, by setting $\kappa=0$. Speaker location tracking can be jointly performed with diarisation within the SSPF model, by using location features in the form of either the DOA with an emission likelihood of \eqref{eq:doa_emission}, or the SSL with an emission likelihood of \eqref{eq:ssl_emission}. A comparison of these features on the \emph{dev} set is shown in Table \ref{tab:features}. The results suggest that both DOA and SSL features may yield small gains over using only $d$-vectors, thereby suggesting that jointly performing speaker tracking with clustering may aid in the diarisation task. The results also agree with \cite{wong2021} in suggesting that location features may be complementary to speaker embeddings for diarisation. SSL features do not show any significant gain over DOA features. In the remaining experiments, the SSL features were used.

\begin{table}[t]
\centering
\caption{Location observation feature type}
\label{tab:features}
\begin{tabular}{l|c}
\hline
Observations&\emph{dev} speaker-attributed WER (\%)\\
\hline\hline
$d$-vector&17.65\\
$d$-vector + DOA&17.56\\
$d$-vector + SSL&17.55\\
\hline
\end{tabular}
\end{table}

\begin{table}[h]
\centering
\caption{Posterior aggregation methods}
\label{tab:decoding_method}
\begin{tabular}{c|c}
\hline
Aggregation method&\emph{dev} speaker-attributed WER (\%)\\
\hline\hline
sum&17.55\\
product&17.56\\
majority voting&17.56\\
\hline
\end{tabular}
\end{table}

As is described in Section \ref{sec:decoding}, the speaker for each word can be chosen by aggregating the per-frame state posteriors within each word using either a sum, product, or majority voting. Table \ref{tab:decoding_method} assesses these aggregation techniques on the \emph{dev} set. There does not seem to be any significant difference between the performances of these three aggregation methods.

\begin{table}[h]
\centering
\caption{Restricting speaker transitions to word boundaries}
\label{tab:restricted_decoding}
\begin{tabular}{cc|cc}
\hline
\multicolumn{2}{c|}{Restrict in}&\multicolumn{2}{c}{\emph{dev} speaker-attributed WER (\%)}\\
forward&backward&forward&backward\\
\hline\hline
no&no&17.55&17.72\\
yes&no&17.60&17.78\\
yes&yes&17.65&17.77\\
\hline
\end{tabular}
\end{table}

Section \ref{sec:decoding} also describes the possibility of restricting the speaker transitions, such that speaker changes are only allowed at word boundaries. This restriction can be applied in either or both of the forward and backward passes. Table \ref{tab:restricted_decoding} assess these restrictions on the \emph{dev} set. Here, the backward pass used only 1000 sub-sampled particles for faster experimentation. The results suggest that there may not be any significant gain yielded by enforcing these restrictions. The 17.60\% and 17.65\% forward pass speaker-attributed WERs for when speaker transitions are restricted in the forward pass differ because of the stochasticity of the SSPF model.

\begin{table}[t]
\centering
\caption{Effect of explicitly modelling movement}
\label{tab:sspf_vs_hmm}
\begin{tabular}{c|c|ccc}
\hline
&&\multicolumn{3}{c}{Speaker-attributed WER (\%)}\\
Test set&Model&stationary&moving&average\\
\hline\hline
\multirow{2}{*}{\emph{dev}}&HMM&16.59&18.19&17.53\\
&SSPF&16.68&18.14&17.55\\
\hline
\multirow{2}{*}{\emph{eval}}&HMM&19.45&15.26&16.02\\
&SSPF&19.54&15.17&16.00\\
\hline
\end{tabular}
\end{table}

Table \ref{tab:sspf_vs_hmm} compares the SSPF against the baseline HMM from \cite{wong2021}, on both the \emph{dev} and \emph{eval} sets. Here the meetings were categorised into those with and without speaker movements. A meeting was considered to have movement if that meeting had at least one speaker, such that it was possible to find two disjoint angular arcs of at least $\frac{\pi}{6}$ radians each, and that speaker spent at least 30s of active speech in each of the two remaining regions that were not covered by these two arcs, based on manually transcribed location information from video data. The results suggest that the SSPF may improve the speaker-attributed WER performance over the HMM for meetings that have movement. Although the improvements may be small for each of the \emph{dev} and \emph{eval} sets, the improvements are consistent across both data sets. However, the SSPF seems to degrade the performance of stationary meetings compared to the HMM. If speakers are fairly stationary through a meeting, then their static location information may be particularly useful for the diarisation task. This scenario may fit particularly well with the assumptions of the HMM, which does not explicitly model temporal changes in the speaker locations. It is shown in \cite{wong2021} that expectation-maximisation fine-tuning of the initial state and state transition probabilities on the current test meeting yield improvements for the HMM. It is difficult to perform per-meeting fine-tuning of the analogous parameters in the SSPF in a computationally feasible manner, and these parameters were instead only initialised from the AHC hypothesis. Despite this, the SSPF is able to perform comparably with the HMM on average.

\begin{figure}[h]
\centering
\includegraphics[width=0.48\textwidth]{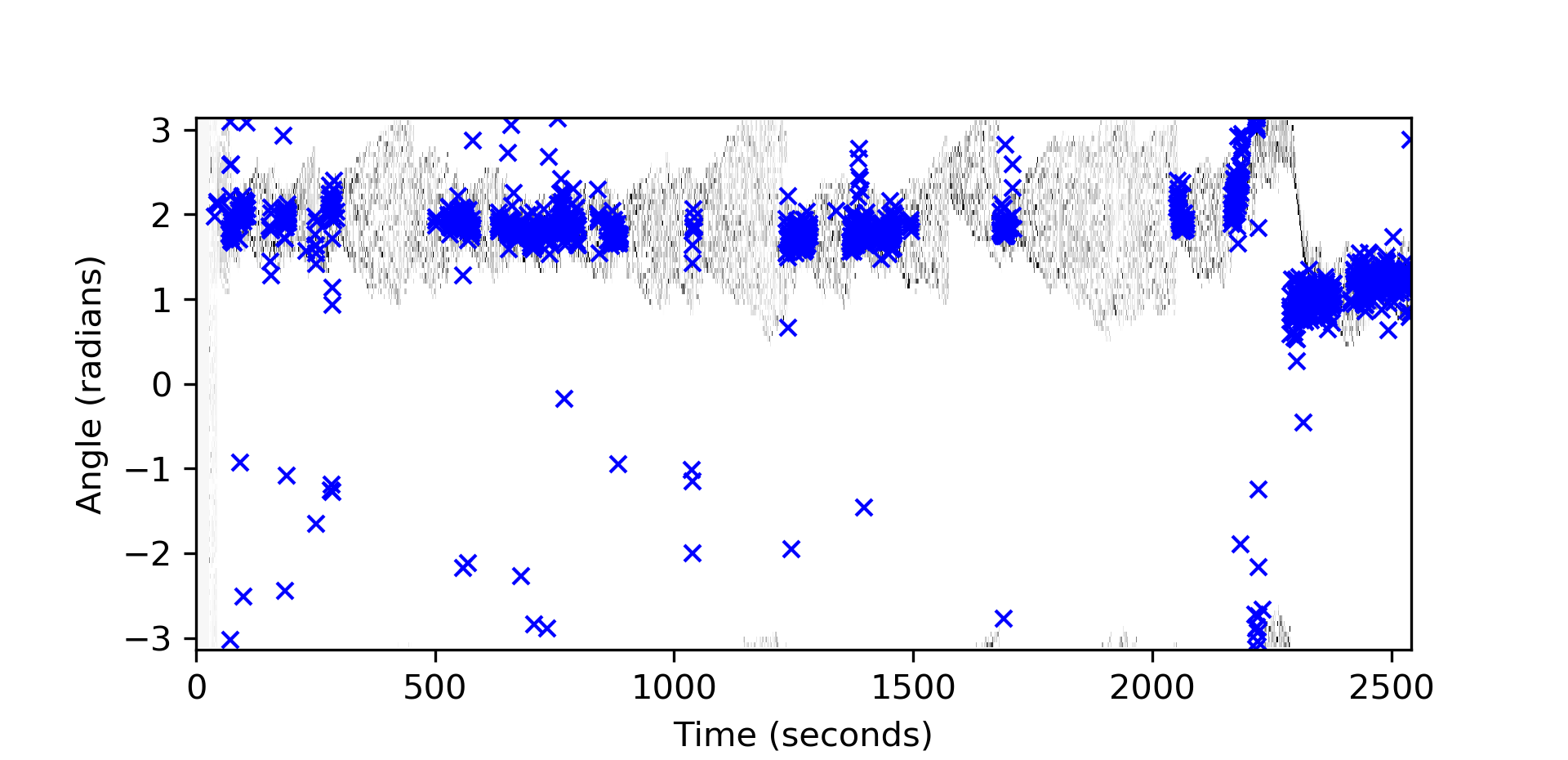}
\caption{Example prediction of a speaker's location. Blue crosses represent the DOA observations, while the heat map shows the weighted distribution of the particles, where darker means higher probability}
\label{fig:trace}
\end{figure}

An advantage of the SSPF over the HMM is that the SSPF can yield the estimated locations of each of the speakers as they move, through the duration of the meeting. An example of such a predicted location trace after the forward pass is illustrated in Figure \ref{fig:trace}. The location estimation continues, even when the speaker is silent. The particles express growing uncertainty about the speaker's location, as the duration of silence increases. This predicted location information may be useful to downstream tasks.

\section{Conclusion}

This paper has proposed a framework to jointly perform diarisation and speaker location tracking. A switching state-space model is implemented as a particle filter, with discrete chains that represent speaker turns, which are used to switch between continuous chains that express speaker locations. This model is shown to perform comparably with a previously proposed HMM diarisation approach that models static speaker locations.

\bibliographystyle{IEEEbib}
\bibliography{strings,refs}

\end{document}